\documentclass[floatfix,prd,superscriptaddress,nofootinbib,amsmath,amssymb,aps,twocolumn]{revtex4-1}
\usepackage{graphicx}
\usepackage{dcolumn}
\usepackage{bm}
\usepackage{mathrsfs}
\usepackage{xcolor,graphicx}
\usepackage{dcolumn}
\usepackage{bm}
\usepackage{enumerate}
\usepackage{feynmf}
\usepackage{subfigure}
\usepackage{todonotes}
\usepackage{multirow}
\usepackage{hyperref}
\usepackage{xspace}
\usepackage{float}
\usepackage{ulem}
\usepackage{diagbox}

\newcommand{\nn}{\nonumber}

\allowdisplaybreaks[4]

\begin{document}

\title{Probing CP Violation through Vector Boson Fusion at High-Energy Muon Colliders}

    \author{Qing-Hong Cao}
    \email{qinghongcao@pku.edu.cn }
    \affiliation{School of Physics, Peking University, Beijing 100871, China}
     \affiliation{School of Physics, Zhengzhou University, Zhengzhou 450001, China}
    \affiliation{Center for High Energy Physics, Peking University, Beijing 100871, China}

\author{Jian-Nan Ding}
\email{dingjn23@pku.edu.cn}
\affiliation{Center for High Energy Physics, Peking University, Beijing 100871, China}

   \author{Yandong Liu}
    \email{ydliu@bnu.edu.cn} 
    \affiliation{School of Physics and Astronomy, Beijing Normal University, Beijing, 100875, China}
    \affiliation{Key Laboratory of Multiscale Spin Physics (Ministry of Education), Beijing Normal University, Beijing, 100875, China}

\author{Jin-Long Yuan}
\email{jinlongyuan@pku.edu.cn}
\affiliation{School of Physics, Peking University, Beijing 100871, China}

\begin{abstract}
We investigate CP-violating effects in electroweak interactions at future high-energy muon colliders within the Standard Model Effective Field Theory (SMEFT) framework. Focusing on four dimension-six CP-odd operators---$ \mathcal{O}_{\widetilde{W}},  \mathcal{O}_{H\widetilde{W}},  \mathcal{O}_{H\widetilde{W}B},  \mathcal{O}_{H\widetilde{B}}$---we analyze vector boson fusion production of $W$ and Higgs bosons using CP-odd observables and their asymmetries. With detailed simulations including parton showering, hadronization, and detector effects, we derive exclusion sensitivities through a binned likelihood analysis. For example, at $\sqrt{s} = 3$~TeV with 2~ab$^{-1}$, the coefficient $C_{\widetilde{W}}$ can be constrained at the $\mathcal{O}(0.02)$ level, improving to $\mathcal{O}(0.008)$ at 10~TeV with 2~ab$^{-1}$, and $\mathcal{O}(0.003)$ with 10~ab$^{-1}$. These results significantly surpass current LHC and projected ILC sensitivities, demonstrating the unique potential of high-energy muon colliders to provide direct and model-independent probes of CP violation in the electroweak sector.
\end{abstract}

\maketitle

\section{Introduction} \label{Sec:Introduction}

The combined Charge-Parity (CP) symmetry is a cornerstone of modern particle physics. 
Its violation, first observed in neutral kaon decays~\cite{Christenson:1964fg}, revealed that nature does not universally respect this symmetry. 
CP violation is not merely an exotic feature of the weak interaction, but plays a fundamental role in cosmology: it is one of the three Sakharov conditions required to generate the observed baryon asymmetry of the Universe (BAU)~\cite{Sakharov:1967dj}. 
The Standard Model (SM) contains a CP-violating source through the Cabibbo-Kobayashi-Maskawa (CKM) phase~\cite{Cabibbo:1963yz,Kobayashi:1973fv}, yet this contribution is far too small to account for the BAU~\cite{Cohen:1993qsc,Cohen:1993nk,Davidson:2008bu}. 
This discrepancy strongly motivates the search for new sources of CP violation beyond the SM (BSM)~\cite{Lee:1973iz,Lee:1974jb,Mohapatra:1974hk,Pilaftsis:1997jf,Covi:1996wh,Affleck:1984fy,Fukugita:1986hr,Akhmedov:1998qx,Luty:1992un,Nardi:2006fx,Buchmuller:2005eh,Morrissey:2012db,Zurek:2013wia,Beneke:2002ks}.

A primary strategy in the search for BSM physics has been the direct production of new particles with CP-violating interactions at high-energy colliders. 
Direct searches at the Large Hadron Collider (LHC) have yet to discover evidence of such new physics, suggesting that the relevant mass scales may lie beyond its direct reach. In this regime, the Standard Model Effective Field Theory (SMEFT) provides a systematic and model-independent framework to parameterize low-energy manifestations of heavy new dynamics~\cite{Grzadkowski:2010es,Giudice:2007fh,Masso:2012eq}. Higher-dimensional operators encode the effects of heavy states, with CP violation arising either from inherently CP-odd Hermitian operators with real Wilson coefficients, or from non-Hermitian operators with complex coefficients introducing explicit CP phases~\cite{Alonso:2013hga,Degrande:2021zpv}. Experimental measurements can thus constrain broad classes of BSM scenarios through SMEFT operators, without relying on detailed ultraviolet completions.

Empirical probes of new CP violation follow two complementary strategies: (1) Low-energy precision observables, such as electric dipole moments (EDMs), yield extremely strong but inclusive bounds: a non-zero EDM would reflect the cumulative effect of many operators, obscuring the origin of CP violation~\cite{ACME:2018yjb,Panico:2018hal}; (2) High-energy colliders, by contrast, provide exclusive probes: by reconstructing the full kinematics of selected final states, they can isolate contributions from specific operators~\cite{Biekotter:2021int,Esmail:2024gdc,Hall:2022bme,Asteriadis:2024xuk}. At the LHC, CP-odd effects have been studied in vector-boson and Higgs production~\cite{ATLAS:2020nzk,CMS:2021nnc,ATLAS:2018hxb,ATLAS:2022fnp}, but the achievable precision is limited by hadronic uncertainties even at the HL-LHC~\cite{Degrande:2021zpv}. Future lepton colliders—particularly high-energy muon colliders—are expected to deliver dramatic improvements in sensitivity, both to specific new physics models and to higher-dimensional operators, through precise cross-section measurements~\cite{deBlas:2022ofj,Gurkanli:2024qaf,Costantini:2020stv,Han:2022edd} and through differential distributions that probe effective vertices~\cite{Ruhdorfer:2024dgz}.
A multi-TeV muon collider effectively acts as a ``gauge boson collider", where vector boson fusion (VBF) and scattering dominate, enabling precise studies of the electroweak sector.

\begin{figure}[t]
\centering
\includegraphics[scale=0.27]
{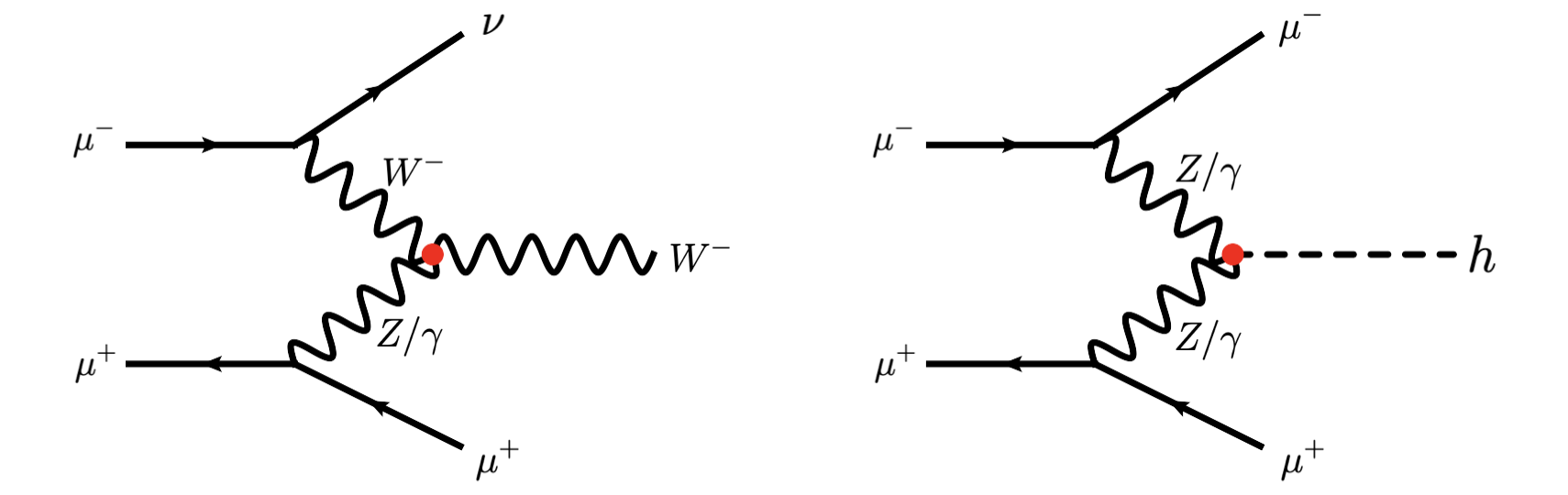}
\caption{\label{fig:FeynmanDiagram} Representative Feynman diagrams for W-boson (left) and Higgs-boson (right) production via vector boson fusion (VBF). The red dots mark the vertices induced by CP-odd operators.}
\end{figure}

In this work we investigate CP violation in electroweak interactions at a high-energy muon collider within the SMEFT framework. We focus on four Hermitian, dimension-six CP-odd operators involving the Higgs doublet and the electroweak gauge fields:
\begin{equation}\label{eq:Relevant_CPodd_operators}
\begin{aligned}
 \mathcal{O}_{\widetilde{W}} &= \epsilon_{IJK} \widetilde{W}^{I\,\nu}_{\mu} W^{J\,\rho}_{\nu} W^{K\,\mu}_{\rho}, \\
 \mathcal{O}_{H\widetilde{W}} &= (H^\dagger H) \,\widetilde{W}^I_{\mu\nu} W^{I\,\mu\nu}, \\
 \mathcal{O}_{H\widetilde{W}B} &= (H^\dagger \tau^I H)\,\widetilde{W}^I_{\mu\nu} B^{\mu\nu}, \\
 \mathcal{O}_{H\widetilde{B}} &= (H^\dagger H)\,\widetilde{B}_{\mu\nu} B^{\mu\nu},
\end{aligned}
\end{equation}
where $H$ is the Higgs doublet, $W^I_{\mu\nu}$ and $B_{\mu\nu}$ are the $SU(2)_L$ and $U(1)_Y$ field strengths, $\tau^I$ are the Pauli matrices, and $\widetilde{X}_{\mu\nu} \equiv \tfrac{1}{2}\varepsilon_{\mu\nu\rho\sigma} X^{\rho\sigma}$ denotes the dual tensor. We analyze their effects in $W$- and Higgs-boson production via VBF, focusing on the processes $\mu^+\mu^-\!\to \mu^\pm \nu W^\mp (\to jj)$ and $\mu^+\mu^-\!\to \mu^+\mu^- h (\to b\bar{b})$. The corresponding Feynman diagrams are illustrated in FIG.\,\ref{fig:FeynmanDiagram}.

\section{CP-Odd Observables in Electroweak Boson Production}\label{sec:CPoddVariable}

\begin{table}
\centering
\caption{\label{cs:Wjj}Cross sections ($\sigma $, fb) from the interference between high-dimension operators and the SM, with $\epsilon$ defined in Eq.~\ref{eq:epsilon:gaugeboson} to be larger than 0 for the process $\mu^+\mu^- \to  \mu^\pm \nu W^\mp( \to jj )$. The Wilson coefficients are set to 1 and the cut-off scale $\Lambda$ to 1 TeV. }
\resizebox{0.6\columnwidth}{!}{
	\begin{tabular}{|c|c|c|}
    \hline
    \hline
\diagbox{$\sqrt{s}$}{operator} &$\mathcal{O}_{\widetilde{W}} $& $\mathcal{O}_{H\widetilde{W}B} $\\
\hline
3 TeV &85.28 &5.10 \\
\hline
10 TeV & 174.89 & 9.44 \\
\hline
    \hline
    \end{tabular}	
 }
\end{table}

\begin{table}
\centering
\caption{\label{cs:hbb}Cross sections ($\sigma $, fb) from the interference between higher-dimension operators and the SM, with $\epsilon$ defined in Eq.~\ref{eq:epsilon:higgsboson} to be larger than 0 for the process $\mu^+\mu^- \to  \mu^+\mu^- h( \to b\bar{b} )$. The Wilson coefficients are set to 1 and the cut-off scale $\Lambda$ to 1 TeV.  }
\resizebox{0.6\columnwidth}{!}{
	\begin{tabular}{|c|c|c|c|}
    \hline
    \hline
\diagbox{$\sqrt{s}$}{operator} &$\mathcal{O}_{H\widetilde{W}} $& $\mathcal{O}_{H\widetilde{W}B} $ & $\mathcal{O}_{H\widetilde{B}} $ \\
\hline
3 TeV &4.11 &1.25 & 0.194 \\
\hline
10 TeV & 8.32 & 2.47  & 0.67 \\
\hline
    \hline
    \end{tabular}	
 }
\end{table}

The squared matrix element in the presence of SMEFT operators can be expanded as
\begin{equation}
|\mathcal{M}|^2 = |\mathcal{M}_{\rm SM}|^2
+ 2\frac{C_i}{\Lambda^2}\,\mathrm{Re}\!\left(\mathcal{M}_{\rm SM}^* \mathcal{M}_i\right)
+ \frac{C_i^2}{\Lambda^4}|\mathcal{M}_i|^2,
\end{equation}
where $\mathcal{M}_{\rm SM}$ and $\mathcal{M}_i$ are the SM and dimension-six amplitudes, respectively. The quadratic term in $C_i^2/\Lambda^4$ is formally of the same order as dimension-eight interference, and is insensitive to CP-odd phases. We therefore neglect this contribution and retain only the interference term, which carries the leading CP-violating signal.

Because the interference is antisymmetric under CP, its integral over the full phase space vanishes~\cite{Ruhdorfer:2024dgz}. To expose CP-odd effects, one must construct observables sensitive to the antisymmetric structure of the amplitude. A widely used choice is the triple-product correlation
\begin{equation}\label{eq:CPodd_observables}
\epsilon \equiv \hat{z}\cdot (\hat{n}_i\times\hat{n}_j),
\end{equation}
where $\hat{z}$ is the beam axis, and $\hat{n}_i,\hat{n}_j$ are unit vectors along selected final-state momenta. For gauge boson production,
\begin{equation}\label{eq:epsilon:gaugeboson}
\epsilon = \hat{z}\cdot(\hat{n}_\mu\times\hat{n}_\nu),
\end{equation}
with the neutrino reconstructed from momentum conservation, while for Higgs production
\begin{equation}\label{eq:epsilon:higgsboson}
\epsilon = \hat{z}\cdot(\hat{n}_{\mu^+}\times\hat{n}_{\mu^-}).
\end{equation}
These observables measure the sine of the relative azimuthal angle of the final-state leptons, and vanish in the SM up to negligible intrinsic CP violation. We note that, although the neutrino momentum is reconstructed and therefore subject to detector smearing effects, the CP-odd observable defined above depends primarily on the relative orientation between the reconstructed missing momentum and the charged lepton. In particular, the sensitivity arises from the sign of the triple-product correlation, which is significantly more robust against moderate detector smearing than observables relying on the absolute momentum scale. The resulting asymmetry therefore remains a reliable probe of CP-violating effects under realistic experimental conditions.

These CP-odd observables are linearly sensitive to the Wilson coefficients while vanishing in the SM, which provides two important advantages. First, they are not contaminated by contributions from dimension-eight operators at the same order, rendering their interpretation within the dimension-six SMEFT framework theoretically cleaner. Second, within the parameter region where the SMEFT expansion remains valid, we explicitly verify that the sensitivities obtained from these differential observables are stronger than those derived from total rate measurements alone. The linear sensitivity therefore constitutes not only a formal advantage but also leads to a tangible improvement in phenomenological reach.

\begin{figure}[t]
\centering
\includegraphics[scale=0.66]{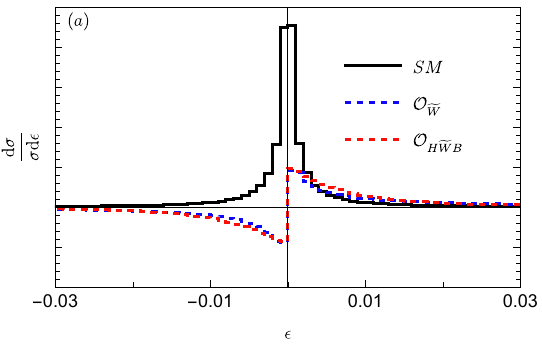}\\
\includegraphics[scale=0.66]{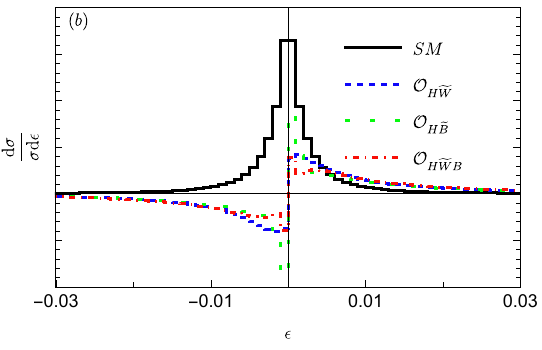}
\caption{\label{fig:dsigma} Normalized distributions of the CP-odd observable $\epsilon$ for (a) $\mu^+\mu^-\!\to \mu^\mp \nu W^\pm$ and (b) $\mu^+\mu^-\!\to \mu^+\mu^-h$ at the parton level. The SM expectation is symmetric, while CP-odd operators induce clear asymmetries.}
\end{figure}

Figure~\ref{fig:dsigma} shows normalized $\epsilon$ distributions for the two processes. The SM yields symmetric spectra, whereas the CP-odd operators generate asymmetric distortions. For $\mu^+\mu^-\!\to \mu^\pm\nu W^\mp$, only $\mathcal{O}_{\widetilde W}$ and $\mathcal{O}_{H\widetilde W B}$ contribute through the $WWZ/\gamma$ vertex, while $\mathcal{O}_{H\widetilde B}$ and $\mathcal{O}_{H\widetilde W}$ are absent due to the anti-symmetric structure of the dual field strength. While in the Higgs boson production, only three Higgs-related operators contribute with varying strength, including destructive interference from $\gamma Z$ exchange.

The interference cross sections from the CP-odd operators are given by
\begin{align} \label{eq:csW}
    &\sigma =\big[  85.28  C_{\widetilde{W}}+5.10  C_{H\widetilde{W}B}  \big]  (\frac{1 \text{TeV}}{\Lambda})^2~\rm fb \nonumber\\
    &\sigma = \big[ 174.89  C_{\widetilde{W}}+9.44  C_{H\widetilde{W}B}  \big] (\frac{1 \text{TeV}}{\Lambda})^2~\rm fb
\end{align}
for the process $\mu^+\mu^-\!\to \mu^\pm \nu W^\mp (\to jj)$ with $\epsilon >0 $ at the 3 TeV and 10 TeV muon collider respectively. 
Similarly, for the process $\mu^+\mu^- \to  \mu^+ \mu^- h( \to b\bar{b} )$, the interference cross sections read
\begin{align} \label{eq:csH}
    &\sigma= \big[ 4.11 C_{H\widetilde{W}}  + 1.25 C_{H\widetilde{W}B} +0.169 C_{H\widetilde{B}} ](\frac{1 \text{TeV}}{\Lambda})^2~\rm fb \nonumber \\
     &\sigma= \big[ 8.32 C_{H\widetilde{W}}  + 2.47 C_{H\widetilde{W}B} +0.67 C_{H\widetilde{B}} ](\frac{1 \text{TeV}}{\Lambda})^2~\rm fb
\end{align}
for center-of-mass energies of 3 TeV and 10 TeV, respectively.
The impact of individual operators on the interference cross sections is summarized in Tables~\ref{cs:Wjj} and~\ref{cs:hbb}. At multi-TeV muon colliders, the dominance of VBF allows the processes to be described as effective gauge-boson scattering. In gauge boson production, $\mathcal{O}_{\widetilde W}$ yields the largest effect, namely about 18 times contribution in comparison with that from operator $\mathcal{O}_{H\widetilde WB}$, due to its complex non-Abelian structure.  While in the Higgs boson production, the three operators contribute comparably but with nontrivial interference patterns.

Finally, we assess the consistency of the SMEFT expansion in the kinematic regime relevant for the observables defined above. Although the collider energy can reach $\sqrt{s}=10$~TeV, the scale controlling the SMEFT expansion is the invariant mass of the hard subprocess, $\sqrt{\hat{s}}$. In the VBF topology, the event kinematics are naturally biased toward lower momentum transfer, with the dominant phase space characterized by $\sqrt{\hat{s}}$ at the TeV scale. The analysis is formulated in terms of the combinations $C_i/\Lambda^2$; for a benchmark cutoff $\Lambda = 1$~TeV, the typical ranges of interest correspond to $C_i \lesssim \mathcal{O}(0.1)$, ensuring that the expansion parameter $C_i \hat{s}/\Lambda^2$ remains perturbative across the relevant phase space. Moreover, the CP-odd observables constructed here are based on asymmetries that are linearly sensitive to the Wilson coefficients, so that the quadratic dimension-six contributions do not affect the consistency of the extraction strategy within the linear SMEFT approximation.

\section{Collider Simulation} \label{sec:collidersim}

In this section, we detail the collider simulation methodology and sensitivity estimation. The four dim-6 operators are implemented using SmeftFR~\cite{Christensen:2008py,Alloul:2013bka,Dedes:2019uzs}, and event samples for both signal and backgrounds, from higher-dimensional operators and SM respectively, are generated using MadEvent~\cite{Alwall:2014hca}. Hadronization processes are simulated with Pythia8~\cite{Sjostrand:2014zea}, and detector effects are incorporated through the Delphes fast simulation framework~\cite{deFavereau:2013fsa}. We use the simulation configuration from the ``MuonCollider" in the Delphes Cards, in which the jet reconstruction algorithm of ``VLCjetR05N2" and b-Jet tagging algorithm of ``MuonColliderDet$\_$BTag$\_$90" are chosen.

\subsection{$W$-boson Production}

For the $W$-boson production process, we focus on the hadronic decays of the $W$ boson to maximize the signal event yield. Consequently, the signal is characterized by the final state $\mu^\pm \nu/\bar{\nu} + 2j$.
The relevant background processes include: (I) vector boson fusion (VBF) production of $W$ bosons, e.g., $\mu^+\mu^-  \to \mu^- \bar{\nu}_\mu W^+$ (or $\mu^+ \nu_\mu W^-$); (II) gauge boson pair production, $\mu^+\mu^-  \to W^\pm W^\mp$, with subsequent semi-leptonic decays; and (III) gauge boson radiation associated with quark pair production, e.g., $\mu^+\mu^-  \to jj W^\pm$, where $j$ denotes a first- or second-generation light quark. Among these, the dominant background arises from the VBF $W$-boson production process. Other background processes are relatively suppressed at high collision energies, scaling as $1/s$, thus becoming less significant at higher-energy muon colliders, such as those operating at 3 TeV or 10 TeV. To suppress contamination from the large background process induced by initial-state photon emission, $a \mu^\pm \to \bar{\nu}_\mu/\nu_\mu W^\mp$, we impose preselection cuts on the leptons in the VBF background process, requiring $|\eta_\ell| < 7$ and $p_T^\ell  > 10~\text{GeV}$.

Event selection is performed using the following basic criteria named as CUT-I:
\begin{align}
&N_\mu=1,\quad p_T^\mu > 10~\text{GeV},\quad |\eta_\mu| < 5.0,\quad \Delta R_{\mu j}>0.4, \nn\\
&N_j=2,\quad p_T^j > 20~\text{GeV}, \quad |\eta_j| < 2.5,\quad \Delta R_{jj}>0.4, \nn
\end{align}
where $N_{\mu,j}$ denote the numbers of muons and jets, $p_T$ represents the transverse momentum, $\eta$ is the pseudo-rapidity, and $\Delta R_{ij}\equiv\sqrt{(\eta_i-\eta_j)^2+(\phi_i-\phi_j)^2}$ corresponds to the angular separation between particles $i$ and $j$. The jet number requirement ($N_j=2$) effectively suppresses reducible backgrounds with additional jets, such as $\mu^+\mu^-\to\mu\nu ZW\to\mu\nu 4j$.

To enhance signal background events ratio, additional optimized selection cuts named CUT-II are applied:
\begin{equation}
m_{\mu\nu}>1000~\text{GeV},\quad |m_{jj}-m_W|<20~\text{GeV},
\end{equation}
where $m_{\mu\nu}$ is the invariant mass of the muon and the neutrino, $m_{jj}$ denotes the dijet invariant mass, and $m_W=80.5$ GeV is the $W$-boson mass. The first optimized cut significantly reduces backgrounds in which the muon and neutrino originate from radiative $W$ bosons, preserving the signal efficiency characteristic of forward-scattering particles in VBF processes~\cite{Rauch:2016pai}. The second cut specifically targets the irreducible background arising from $t$-channel $ZW$ scattering, ensuring the dijet invariant mass aligns closely with the $W$-boson mass.

For the higher-energy scenario at $\sqrt{\hat{s}}=10$ TeV, where forward muons and neutrinos possess significantly higher energies, the optimized invariant mass selection criterion in CUT-II is adjusted to:
\begin{equation}
m_{\mu\nu}>3000~\text{GeV},\quad |m_{jj}-m_W|<20~\text{GeV},
\end{equation}
to maximize the signal event and remove the backgrounds events.

The cut flow for  signal and backgrounds are shown in Tab.~\ref{tab:W3tev} and Tab.~\ref{tab:W10tev}. It is clearly shown that the CUT-II, namely $m_{\mu\nu}$ characterizing the VBF process and $W$-boson reconstruction, is very efficient to abandon the backgrounds event and enhance the signal backgrounds events ratios. 

\begin{table}
\centering
\caption{\label{tab:W3tev} 
The cut flows of the signal ($\mu^+\mu^-\to \mu^\pm \nu  jj$) with $\epsilon>0$, benchmark Wilson coefficient $C_i=1$, $\Lambda=1$ TeV and SM backgrounds cross sections (in fb) at a 3 TeV muon collider.}
	\begin{tabular}{|c|c|c|c|}
    \hline
    \hline
\diagbox{$\sigma (\mathrm{fb})$}{CUTs}&pre-selections& CUT-I & CUT-II \\
    \hline
$\sigma_{W^\pm(\to jj) \mu^\mp \nu } (\mathcal{O}_{\widetilde{W}})$ & 85.28& 80.88&30.94 \\
    \hline
$\sigma_{W^\pm(\to jj) \mu^\mp \nu } (\mathcal{O}_{H\widetilde{W}B})$ & 5.10& 4.72 &1.02 \\
\hline
$\sigma_{W^\pm \mu^\mp \nu} $ & 4705.8 &4341.2&  758.63\\
\hline
$\sigma_{W^\pm W^\mp (\to \mu^\pm \nu jj)}$ &66.42 &41.67 &0.459 \\
\hline
$\sigma_{jjW^\pm(\to \mu^\pm \nu)}$ &5.62 & 5.33&0.00444 \\
\hline
    \hline
    \end{tabular}	
\end{table}

\begin{table}
\centering
\caption{\label{tab:W10tev}The cut flows of the signal ($\mu^+\mu^-\to \mu^\pm \nu  jj$) with $\epsilon>0$, benchmark Wilson coefficient $C_i=1$, $\Lambda=1$ TeV and SM backgrounds cross sections (in fb) at a 10 TeV muon collider.}
	\begin{tabular}{|c|c|c|c|}
    \hline
    \hline
\diagbox{$\sigma (\mathrm{fb})$}{CUTs}&pre-selections& CUT-I & CUT-II \\
    \hline
$\sigma_{W^\pm(\to jj) \mu^\mp \nu } (\mathcal{O}_{\widetilde{W}})$ & 174.89&157.91 &54.03 \\
    \hline
$\sigma_{W^\pm(\to jj) \mu^\mp \nu } (\mathcal{O}_{H\widetilde{W}B})$ &9.44 & 7.58 & 0.86\\
\hline
$\sigma_{W^\pm \mu^\mp \nu} $ & 7779.34 & 6327.68& 338.69 \\
\hline
$\sigma_{W^\pm W^\mp (\to \mu^\pm \nu jj)}$ & 8.39 &2.29 & 0.075 \\
\hline
$\sigma_{jjW^\pm(\to \mu^\pm \nu)}$ &5.93& 5.58 & 4.2$\times 10^{-3}$ \\
\hline
    \hline
    \end{tabular}	
\end{table}

\subsection{Higgs Boson Production}

For the Higgs boson production channel, we require the Higgs to decay into $b\bar{b}$, capitalizing on its large branching fraction and the distinctive signatures provided by $b$-jets at lepton colliders. Consequently, the signal is characterized by a pair of muons accompanied by two $b$-jets. The primary background contributions arise from three sources: (I) gauge boson ($Z$) production associated with a muon pair, including vector boson fusion (VBF) production of $Z$ and s-channel $Z$ radiation processes; (II) gauge boson pair production, e.g., $\mu^+\mu^-  \to ZZ \to \mu^+ \mu^- b \bar{b}$; and (III) VBF Higgs boson production, $\mu^+\mu^-  \to \mu^+\mu^-  h(\to b\bar{b})$. Other potential background processes, such as the associated production of a Higgs boson with a $Z$ boson ($hZ$), are strongly suppressed at high energies, scaling as $1/s$, and thus can be considered negligible.

To effectively reconstruct Higgs production through vector boson fusion (VBF), we apply the following basic selection criteria named CUT-I:
\begin{align}
&N_\mu=2,\quad p_T^\mu > 10~\text{GeV},\quad |\eta_\mu| < 5.0,  \nonumber \\
&N_j=2,\quad p_T^j > 20~\text{GeV}, \quad |\eta_j| < 2.5, \quad N_{{\mathrm{b-jet}}}=1,  \nonumber \\
&\Delta R_{\mu\mu}>0.4,\quad \Delta R_{\mu j}>0.4,\quad \Delta R_{jj}>0.4, \nonumber \\
& | m_{jj} - m_h | < 25~\text{GeV}
\end{align}
where $m_h=125$ GeV is the Higgs boson mass.

Additional optimized selection cuts named CUT-II are implemented:
\begin{equation}
m_{\mu\mu} > 500~\text{GeV},\quad \Delta\eta_{\mu\mu}>3.0, \nn
\end{equation}
where $\Delta \eta_{\mu\mu}=|\eta_{\mu^+}-\eta_{\mu^-}|$ represents the pseudo-rapidity difference between the muon and anti-muon pair. The requirement of a large invariant mass and significant pseudo-rapidity separation of the muon pair effectively captures the forward-scattering nature characteristic of VBF processes~\cite{Rauch:2016pai}, substantially suppressing SM background contributions from radiative diagrams. Additionally, requiring the dijet invariant mass to be close to the Higgs boson mass significantly reduces backgrounds involving $b$-jets from $Z$ boson decays or $t$-channel $ZZ$ scattering.

For the higher-energy scenario at $\sqrt{\hat{s}}=10$ TeV, where the final-state muons become more forward and energetic, we adjust the optimized selection criteria  in CUT-II accordingly:
\begin{equation}
m_{\mu\mu}>4000~\text{GeV},\quad \Delta\eta_{\mu\mu}>6.0.
\end{equation}

Table~\ref{tab:h3tev} and ~\ref{tab:h10tev} show the cut flows of the signals and backgrounds. From the tables it is clearly shown that the CUT-I including the Higgs boson reconstruction is more efficient to enhance the signal backgrounds events ratios.

\begin{table}
\centering
\caption{\label{tab:h3tev}The cut flows of the signal ($\mu^+\mu^-\to \mu^+ \mu^-  b\bar{b}$) with $\epsilon>0$, benchmark Wilson coefficient $C_i=1$, $\Lambda=1$ TeV and SM backgrounds cross sections (in fb) at a 3 TeV muon collider.}
	\begin{tabular}{|c|c|c|c|}
    \hline
    \hline
\diagbox{$\sigma (\mathrm{fb})$}{CUTs}&pre-selections& CUT-I & CUT-II \\
    \hline
$\sigma_{\mu^\mp \mu^\mp b\bar{b} } (\mathcal{O}_{H\widetilde{W}} )$ &4.11 &1.24 &1.19 \\
\hline
$\sigma_{\mu^\mp \mu^\mp b\bar{b} } (\mathcal{O}_{H\widetilde{W}B} )$ &1.25 &0.38 &0.36\\
\hline
$\sigma_{\mu^\mp \mu^\mp b\bar{b} } (\mathcal{O}_{H\widetilde{B}} )$ &0.194 &0.0579 &0.0555 \\
\hline
$\sigma_{\mu^+ \mu^- Z(\to b\bar{b}/ c\bar{c})} $ &106.4  &0.136&  0.086\\
\hline
$\sigma_{ZZ (\to \mu^+ \mu^- b\bar{b}/c\bar{c})}$ &0.465 &0.003 &$4.6\times10^{-6}$ \\
\hline
$\sigma_{ \mu^+ \mu^- h(\to b\bar{b}) }$ &41.87 & 11.77&11.34 \\
\hline
    \hline
    \end{tabular}	
\end{table}

\begin{table}
\centering
\caption{\label{tab:h10tev}The cut flows of the signal ($\mu^+\mu^-\to \mu^+ \mu^-  b\bar{b}$) with $\epsilon>0$, benchmark Wilson coefficient $C_i=1$, $\Lambda=1$ TeV and SM backgrounds cross sections (in fb) at the 10 TeV muon collider.}
	\begin{tabular}{|c|c|c|c|}
    \hline
    \hline
\diagbox{$\sigma (\mathrm{fb})$}{CUTs}&pre-selections& CUT-I & CUT-II \\
    \hline
$\sigma_{\mu^\mp \mu^\mp b\bar{b} } (\mathcal{O}_{H\widetilde{W}} )$ &8.32 &2.14 &1.97 \\
\hline
$\sigma_{\mu^\mp \mu^\mp b\bar{b} } (\mathcal{O}_{H\widetilde{W}B} )$ & 2.47& 0.63& 0.58\\
\hline
$\sigma_{\mu^\mp \mu^\mp b\bar{b} } (\mathcal{O}_{H\widetilde{B}} )$ & 0.67 & 0.17 & 0.15 \\
\hline
$\sigma_{\mu^+ \mu^- Z(\to b\bar{b}/ c\bar{c})} $ & 111.69 &0.069& 0.02 \\
\hline
$\sigma_{ \mu^+ \mu^- h(\to b\bar{b}) }$ & 70.78 & 15.89& 15.19 \\
\hline
    \hline
    \end{tabular}	
\end{table}

\subsection{Sensitivity Estimation}

Following the application of the selection criteria outlined above, all surviving events are categorized into two bins based on the CP-odd observable: $\epsilon >0$ and $\epsilon <0$. To constrain contributions from dimension-six operators, we define the following likelihood function:
\begin{align}
L_{\text{SM}/\text{NP}} = \prod_{\epsilon}\frac{\left(n^{\epsilon}_{\text{SM}/\text{NP}}\right)^{n^{\epsilon}_{\text{obs}}} e^{-n^{\epsilon}_{\text{SM}/\text{NP}}}}{n^{\epsilon}_{\text{obs}}!},
\end{align}
where the superscript $\epsilon$ indicates the bin category, either $\epsilon<0$ or $\epsilon>0$. Here, $n^{\epsilon}_{\text{SM}/\text{NP}}$ represents the predicted number of events according to theoretical models (SM alone or SM combined with dimension-six operator contributions) in each respective bin, while $n^{\epsilon}_{\text{obs}}$ denotes the observed event number. Assuming observations consistent with the SM predictions ($n^{\epsilon}_{\text{obs}}=n^{\epsilon}_{\text{SM}}$), the confidence level for excluding the presence of higher-dimensional operator contributions is quantified using the test statistic:
\begin{equation} \label{eq:sta}
\begin{aligned}
\Delta\chi^2 &= -2\log\frac{L_{\text{NP}}}{L_{\text{SM}}}  \\
&=\sum_\epsilon 2(n^{\epsilon}_{\text{NP}} -n^{\epsilon}_{\text{SM}} \log \frac{ n^{\epsilon}_{\text{SM}} + n^{\epsilon}_{\text{NP}} }{ n^{\epsilon}_{\text{SM}} }  ).
\end{aligned}
\end{equation}
For the purpose of this analysis, we present $95\%$ confidence-level exclusion limits corresponding to $\Delta\chi^2 = 4$.
Equation~\ref{eq:sta} assumes purely statistical uncertainties. To assess the impact of systematic effects, we extend the likelihood by introducing a background normalization uncertainty of $X\%$ in each bin, implemented through a nuisance parameter on the SM expectation,
$n_{\rm SM}^{\epsilon} \to n_{\rm SM}^{\epsilon}(1+X\%)$, with a Gaussian prior of width $X\%$, and profile over $n_{\rm SM}^{\epsilon}X\%$ in the test statistic~\cite{Cowan:2010js}.

\section{Result and Discussion}\label{sec:RD}
With all event selection efficiencies for both signals and backgrounds taken into account, we perform a binned likelihood analysis based on the CP-odd observable $\epsilon$, following the procedure outlined above. The resulting $95\%$ confidence-level (C.L.) exclusion limits, corresponding to $\Delta\chi^2=4$ for a single parameter of interest, are summarized in Tab.\ref{tab:Constraints_VBF_gauge} and Tab.\ref{tab:Constraints_VBF_Higgs}. As shown, the Wilson coefficient $C_{\widetilde{W}}$ can be constrained to the order of $0.02$ at a 3 TeV muon collider with an integrated luminosity of 2 ab$^{-1}$. At a 10 TeV muon collider, the sensitivity improves significantly, reaching the order of $0.008$ with 2 ab$^{-1}$ and $0.003$ with 10 ab$^{-1}$. For the coefficient $C_{H\widetilde{W}B}$, the corresponding bounds are weaker: it is constrained to the order of $0.6$ at 3 TeV and $0.48$ at 10 TeV, both with 2 ab$^{-1}$, while with 10 ab$^{-1}$ at 10 TeV the limit improves to about $0.2$. Notably, the constraint on $C_{\widetilde{W}}$ is about 30 times tighter than that on $C_{H\widetilde{W}B}$ at 3 TeV, and about 70 times tighter at 10 TeV. This difference originates from the distinct cut efficiencies (see Tab.\ref{tab:W3tev} and Tab.\ref{tab:W10tev}), which reflect the underlying kinematics of the operators. In particular, $\mathcal{O}_{H\widetilde{W}B}$ receives sizable contributions from the $WW\gamma$ vertex, leading to more forward-peaked final-state particles in vector-boson-fusion processes at high energies, thus reducing the cut acceptance. For Higgs boson production, the three operators $\mathcal{O}_{H\widetilde{W}}$, $\mathcal{O}_{H\widetilde{W}B}$, and $\mathcal{O}_{H\widetilde{B}}$ only interfere with the SM $ZZh$ interaction. Consequently, their cut efficiencies are nearly identical, and the resulting bounds on the corresponding Wilson coefficients $C_{H\widetilde{W}}$, $C_{H\widetilde{W}B}$, and $C_{H\widetilde{B}}$ depend primarily on the collider energy and integrated luminosity; see Tab.~\ref{tab:Constraints_VBF_Higgs}.

To assess the robustness of our projections against experimental systematics, we further evaluate the impact of a background normalization uncertainty modeled as a fractional uncertainty $X\%$ on the SM expectation in each bin. In addition to the purely statistical case $X=0$, we consider two representative benchmark choices, X=10 and X=20, which roughly correspond to optimistic and conservative assumptions commonly adopted in future collider sensitivity studies. We find that even for the conservative choice $X=20$, the exclusion limits degrade by only about $10\%$; see Tab.~\ref{tab:C_VBF_gauge_unc} and \ref{tab:C_VBF_Higgs_unc}. This demonstrates that the projected sensitivities are not driven by overly idealized assumptions and remain stable in the presence of realistic experimental uncertainties.

For comparison, we include existing limits on these operators from current ATLAS measurements~\cite{ATLAS:2020nzk}, the HL-LHC projections~\cite{Degrande:2021zpv}, the ILC~\cite{deBlas:2022ofj}, and low-energy precision observables such as the electron EDM (eEDM)~\cite{Panico:2018hal,ACME:2018yjb}. Note that the ILC constraints are quoted at the $1\sigma$ confidence level. It is evident that a 10~TeV muon collider can provide highly competitive and, in many cases, substantially stronger bounds than those expected from the LHC and ILC; see Tab.~\ref{tab:ConstraintsOnWCs}. While the eEDM measurements yield much more stringent constraints on individual operators, in realistic scenarios, multiple operators may contribute simultaneously, making it impossible to disentangle their individual effects
\begin{equation}
\begin{aligned}
   d_e\propto&(9.77 C_{\widetilde{W}} + 40.76 C_{H\widetilde{W}} -145.23 C_{H\widetilde{W}B} + 122.28 C_{H\widetilde{B}} )  \\
    &\times (\frac{1 \text{TeV}}{ \Lambda})^2.
\end{aligned}
\end{equation}
When compared with Eqs.~\ref{eq:csW} and \ref{eq:csH}, this relation highlights the complementary role of high-energy colliders, which probe different linear combinations of CP-violating operators through distinct correlation patterns among the Wilson coefficients, thereby providing information that is inaccessible to low-energy observables. 
For instance, cancellations among operator contributions can occur in the eEDM (e.g., between $\mathcal{O}_{H\widetilde{W}B}$ and other operators), while the same parameter combinations typically contribute constructively to the CP-odd observables at a muon collider; see Eqs.~\ref{eq:csW} and \ref{eq:csH}.
We note that a direct comparison of constraints obtained at different experiments is nontrivial, as it requires a consistent treatment of renormalization-group running and operator mixing effects across the relevant energy scales, which is beyond the scope of the present work.

\begin{table}
\centering
\caption{The constraints on Wilson coefficients at $95\%$ C.L. in $W$-boson production via vector boson fusion, where the cutoff scale of new physics is normalized to $\Lambda=1$ TeV.} \label{tab:Constraints_VBF_gauge}
	\begin{tabular}{c|c|c|c}
    \hline
    \hline
     Wilson  &$\sqrt{\hat{s}}=3$ TeV & $\sqrt{\hat{s}}=10$ TeV & $\sqrt{\hat{s}}=10$ TeV \\
     Coefficients &$\mathcal{L}=2$ ab$^{-1}$ &$\mathcal{L}=2$ ab$^{-1}$ &$\mathcal{L}=10$ ab$^{-1}$  \\
    \hline
    $C_{\widetilde{W}}$ &$[-0.02,0.02]$ &$[-0.0076,0.0076]$ &$[-0.0034,0.0034]$   \\
    $C_{H\widetilde{W}B}$   &$[-0.6,0.6]$  &$[-0.48,0.48]$ &$[-0.21,0.21]$    \\
    \hline
    \hline
    \end{tabular}	
\end{table}

\begin{table}
\centering
\caption{The constraints on Wilson coefficients at $95\%$ C.L. in Higgs production via vector boson fusion, where the cutoff scale of new physics is normalized to $\Lambda=1$ TeV. } \label{tab:Constraints_VBF_Higgs}
	\begin{tabular}{c|c|c|c}
    \hline
    \hline
     Wilson &$\sqrt{\hat{s}}=3$ TeV, &$\sqrt{\hat{s}}=10$ TeV, & $\sqrt{\hat{s}}=10$ TeV,  \\
     Coefficients &$\mathcal{L}=2$ ab$^{-1}$ &$\mathcal{L}=2$ ab$^{-1}$ &$\mathcal{L}=10$ ab$^{-1}$  \\
    \hline
    $C_{H\widetilde{W}}$    &$[-0.064,0.064]$ &$[-0.044,0.044]$ &$[-0.02,0.02]$  \\
    $C_{H\widetilde{W}B}$  &$[-0.21,0.21]$  &$[-0.15,0.15]$ &$[-0.067,0.067]$   \\
    $C_{H\widetilde{B}}$   &$[-1.36,1.36]$ &$[-0.58,0.58]$  &$[-0.26,0.26]$    \\
    \hline
    \hline
    \end{tabular}	
\end{table}

\begin{table}[]
    \centering
    \caption{Constraints on the Wilson coefficients at the 10 TeV muon collider with an integrated luminosity of 10 ab$^{-1}$, including a background normalization uncertainty of $X\%$ in the test statistic. The bounds are obtained at $95\%$ C.L. in $W$-boson production via vector boson fusion, with the cutoff scale of new physics normalized to $\Lambda = 1$ TeV.}
    \label{tab:C_VBF_gauge_unc}
    \begin{tabular}{c|c|c|c}
    \hline\hline
         & $X=0$  & $X=10$  & $X=20$\\ 
         \hline
      $C_{\widetilde{W}}$    & $[-0.0034,0.0034]$  & $[-0.0036,0.0036]$  & $[-0.0037,0.0037]$ \\
      $C_{\widetilde{W}B}$    &  $[-0.21,0.21]$ & $[-0.22,0.22]$ & $[-0.23,0.23]$\\
      \hline
      \hline
     \end{tabular}
\end{table}

\begin{table}[]
    \centering
    \caption{Constraints on the Wilson coefficients at the 10 TeV muon collider with an integrated luminosity of 10 ab$^{-1}$, including a background normalization uncertainty of $X\%$ in the test statistic. The bounds are obtained at $95\%$ C.L. in Higgs production via vector boson fusion, with the cutoff scale of new physics normalized to $\Lambda = 1$ TeV.}
    \label{tab:C_VBF_Higgs_unc}
    \begin{tabular}{c|c|c|c}
    \hline\hline
         & $X=0$  & $X=10$  & $X=20$\\ 
         \hline
     $C_{H\widetilde{W}}$   & $[-0.02,0.02]$  & $[-0.021,0.021]$  & $[-0.022,0.022]$ \\
      $C_{H\widetilde{W}B}$   &  $[-0.067,0.067]$ & $[-0.07,0.07]$ & $[-0.074,0.074]$\\
    $C_{H\widetilde{B}}$   & $[-0.26,0.26]$ & $[-0.27,0.27]$  &  $[-0.28,0.28]$  \\
      \hline
      \hline
     \end{tabular}
\end{table}

\begin{table*}
\centering
\caption{Constraints on the CP-odd operators from different processes in current and future projections, with the cutoff scale of new physics normalized to $\Lambda = 1$~TeV. Unless otherwise specified, the bounds correspond to the $2\sigma$ confidence level, except for the ILC results, which are quoted at $1\sigma$.}\label{tab:ConstraintsOnWCs}
\resizebox{0.75\textwidth}{!}{
	\begin{tabular}{c|c|c|c|c|c}
	\hline
        \hline
     Processes &$\sqrt{\hat{s}}$ (TeV) &$\mathcal{O}_{\widetilde{W}}$ &$\mathcal{O}_{H\widetilde{W}}$ &$\mathcal{O}_{H\widetilde{W}B}$ &$\mathcal{O}_{H\widetilde{B}}$ \\
     \hline
     \multirow{2}{*}{this work} &$3.0$ &$[-0.02,0.02]$ &$[-0.064,0.064]$ &$[-0.21,0.21]$ &$[-1.36,1.36]$ \\
      &$10.0$ &$[-0.0034,0.0034]$ &$[-0.02,0.02]$ &$[-0.067, 0.067]$ &$[-0.26,0.26]$\\
      \hline
    ATLAS ($pp\to jjZ$) \cite{ATLAS:2020nzk} &$13.0$ &$[-0.11,0.14]$ &$-$ &$[0.23,2.34]$ &$-$\\
      \hline
    HL-LHC ($pp\to W\gamma$) \cite{Degrande:2021zpv} &$13.0$ &$[-0.15,0.15]$ &$-$ &$[-0.11,0.11]$ &$-$\\
    \hline
    ILC \cite{deBlas:2022ofj} & &$-$ 
    &$[-0.032,0.032]$
    &$[-0.062,0.062]$ &$[-0.169,0.169]$ \\
    \hline
    eEDM \cite{Panico:2018hal,ACME:2018yjb} & &$\leq1.77\times 10^{-4}$ &$\leq4.14\times 10^{-5}$ &$\leq1.16\times 10^{-5}$ &$\leq1.38\times 10^{-5}$ \\
    \hline
    \hline
	\end{tabular}	
}
\end{table*}

\begin{acknowledgments}
 The work of Q.C. J.D. Y.L. and J.Y. is partly supported by the National Science Foundation of China under Grant Nos. 12075257, 12175016 and 12235001 and the National Key R$\&$D Program of China under Grant No. 2023YFA1607104.

\end{acknowledgments}

\bibliographystyle{apsrev}
\bibliography{ref}
\end{document}